# Tracking gas–solid hydration kinetics by non-contact in situ transflection micro-FTIR spectroscopy

Kunihiko Shizume[1*], Hidetoshi Miyazaki[1], Yasushi Hamanaka[1]

[1] Nagoya Institute of Technology

## ABSTRACT

Gas–solid reaction kinetics is important in applications such as thermochemical heat storage (TCHS), but experimental evaluation remains a bottleneck in materials development. This study evaluates non-contact in situ transflection micro-Fourier transform infrared (FTIR) spectroscopy for tracking hydration reactions between water vapor and inorganic solids. Powder layers of $La_2(SO_4)_3$, $CaSO_4$, and CaO, exhibiting different hydration mechanisms and kinetics, were deposited on an Au-coated glass substrate and exposed to humidified $N_2$. Evolution of water- and hydroxyl-related IR absorption band areas was compared with thermogravimetry (TG) derived hydration numbers. The FTIR band-area profiles were broadly consistent with the TG-derived hydration trends, showing close agreement for $La_2(SO_4)_3$ and $CaSO_4$. For CaO, some differences were observed between the FTIR and TG results, possibly because CaO is sensitive to differences in gas-phase conditions, particularly the initial increase in humidity, between the two measurements. The spectra distinguished molecular-water uptake in $La_2(SO_4)_3$ and $CaSO_4$ from hydroxide formation in CaO. They also detected carbonate formation during CaO hydration. Thus, in situ time-resolved FTIR separately monitored individual chemical reactions, whereas TG recorded only their combined effect on the sample mass. These results demonstrate the potential of transflection micro-FTIR for chemically selective monitoring of gas–solid reactions in small localized samples.

## 1. Introduction

The kinetics of gas–solid processes directly affects the performance of a wide range of technologies, including thermochemical energy storage, adsorption-based gas separation, and chemical looping.[1–4] For gas–solid reactions, reaction kinetics is governed by the complex interplay of mass transport within and around the solid, interfacial chemical reactions, and heat transfer.[5–8] The relative importance of these processes depends on the material, reaction progress, and operating conditions,[6,7,9] making reaction kinetics difficult to predict a priori. Experimental evaluation therefore remains essential in materials development.

Among these technologies, thermochemical heat storage (TCHS) provides a particularly clear example of the importance of gas–solid reaction kinetics.[1,10] TCHS stores and releases thermal energy through reversible endothermic and exothermic chemical reactions. Among the various reaction classes considered for TCHS, hydration and dehydration reactions involving the uptake and release of water vapor by inorganic solids have been widely investigated as a means of utilizing low-temperature waste heat.[11,12]

However, hydration or dehydration often proceed too slowly under low-temperature operating conditions, limiting the practical applicability of many candidate materials.[13,14] This motivates the search for additional materials, but the number of potentially suitable inorganic hydrates is large. The expansion of crystal structure databases, together with first-principles calculations, has enabled large-scale computational screening of candidate materials.[15] For example, 265 hydration reactions of compounds listed in a crystal structure database were evaluated in terms of their operating temperatures and energy densities, allowing promising reactions to be identified.[16] More recently, computational screening has been extended to hypothetical materials: the thermodynamic properties of 7,012 hypothetical hydrates were predicted, and promising candidates were identified based on their thermodynamic stability and energy density.[17] However, such calculations do not guarantee that these candidates will undergo hydration or dehydration on practical timescales. Thus, experimental evaluation of their reaction kinetics remains necessary.[14]

Thermogravimetric (TG) analysis is widely used to evaluate hydration and dehydration kinetics because it can detect small mass changes and provide quantitative information on reaction progress.[11,18] However, conventional TG analysis generally requires a separate measurement for each sample. Experimental evaluation of large numbers of candidate materials can therefore become a bottleneck relative to their computational identification, motivating the development of experimental methods that could enable high-throughput evaluation of reaction kinetics.

Time-resolved in situ Fourier transform infrared (FTIR) spectroscopy has been used to determine reaction kinetics in other chemical systems.[19] Molecular water and hydroxyl groups exhibit characteristic vibrational bands in FTIR spectra, making the technique suitable for monitoring hydration reactions. In situ attenuated total reflection FTIR (ATR-FTIR) measurements have followed chemical and phase evolution during the hydration of cement pastes and provided insight into the associated reaction mechanisms.[20,21] More recently, ATR-FTIR responses during cement hydration have been compared with X-ray diffraction and calorimetry. Characteristic integral band areas were found to be sufficient for qualitative tracking of hydrate formation, whereas detailed peak fitting provided additional structural information.[22] These studies employed contact-based ATR measurements of cement pastes. In contrast, transflection provides a non-contact geometry that allows solid samples, including powder layers deposited on a reflective substrate, to be measured. In this geometry, infrared light passes through the sample layer, reflects from the substrate, and passes through the layer again before reaching the detector. However, applying transflection FTIR to powder layers is not straightforward because particle scattering and thickness-dependent optical effects associated with the substrate [23,24] may affect the measured spectra. Therefore, its applicability to time-resolved monitoring of gas–solid hydration kinetics needs to be evaluated.

In this study, we examine the feasibility of using non-contact in situ transflection micro-FTIR spectroscopy to track gas–solid hydration kinetics. Three inorganic materials with relatively high hydration kinetics, corresponding to experimentally manageable reaction timescales, were chosen. $La_2(SO_4)_3$ and $CaSO_4$ were selected as low-temperature TCHS materials exhibiting relatively rapid hydration.[14,25–27] CaO was also included as a practically important TCHS material operating above 500 °C, [1,18] despite not being a low temperature TCHS material. In the present study, CaO was chosen as a representative material exhibiting a longer reaction timescale compared with the other two. The corresponding reversible reactions for TCHS applications are as follows:

$$La_2(SO_4)_3(s) + H_2O(g) \rightleftharpoons La_2(SO_4)_3{\cdot}H_2O(s) \quad (1)$$

$$CaSO_4(s) + 0.5H_2O(g) \rightleftharpoons CaSO_4{\cdot}0.5H_2O(s) \quad (2)$$

$$CaO(s) + H_2O(g) \rightleftharpoons Ca(OH)_2(s) \quad (3)$$

Here, the hydration reactions represented by the left-to-right directions in Eqs. (1)–(3) were investigated using both micro-FTIR and TG. Comparison of the two measurements demonstrated that the time-dependent integrated areas of water- and hydroxyl-related IR absorption bands can be used to evaluate gas–solid hydration kinetics, with results consistent with those obtained by TG.

## 2. Experimental

Figure 1 illustrates the experimental workflow used in this study. Sample powder was deposited on Au-coated glass substrates for in situ micro-FTIR and TG measurements. Hydration was initiated by switching the gas flow from dry to a mixture of dry and humidified $N_2$, generated by passing $N_2$ through a water bubbler. The water vapor pressure, $p_{H_2O}$, at the gas outlet was monitored continuously. The time-dependent hydration behaviors measured by the two techniques were then compared.

## 2.1 Materials

Lanthanum sulfate nonahydrate $La_2(SO_4)_3{\cdot}9H_2O$ (≥ 98.0%, Wako Pure Chemical, Japan), calcium sulfate dihydrate $CaSO_4{\cdot}2H_2O$ (≥ 98.0%, Wako Pure Chemical), and calcium hydroxide $Ca(OH)_2$ (99.9%, High Purity Chemical, Japan) were used as starting materials. All chemical powders were sieved through a 75-μm mesh before use. Each powder was dispersed in ethanol at a concentration of 4 vol%. Aliquot of the resulting dispersion (4 μL for $La_2(SO_4)_3$ and 2 μL for $CaSO_4$ and $Ca(OH)_2$) were deposited onto Au-coated glass substrates with a 100-nm-thick Au film (product no. 1091, Geomatec, Japan) and dried under ambient conditions to form powder layers (Fig. 1(a)). Separate specimens prepared using the same procedure were used for the micro-FTIR and TG measurements.

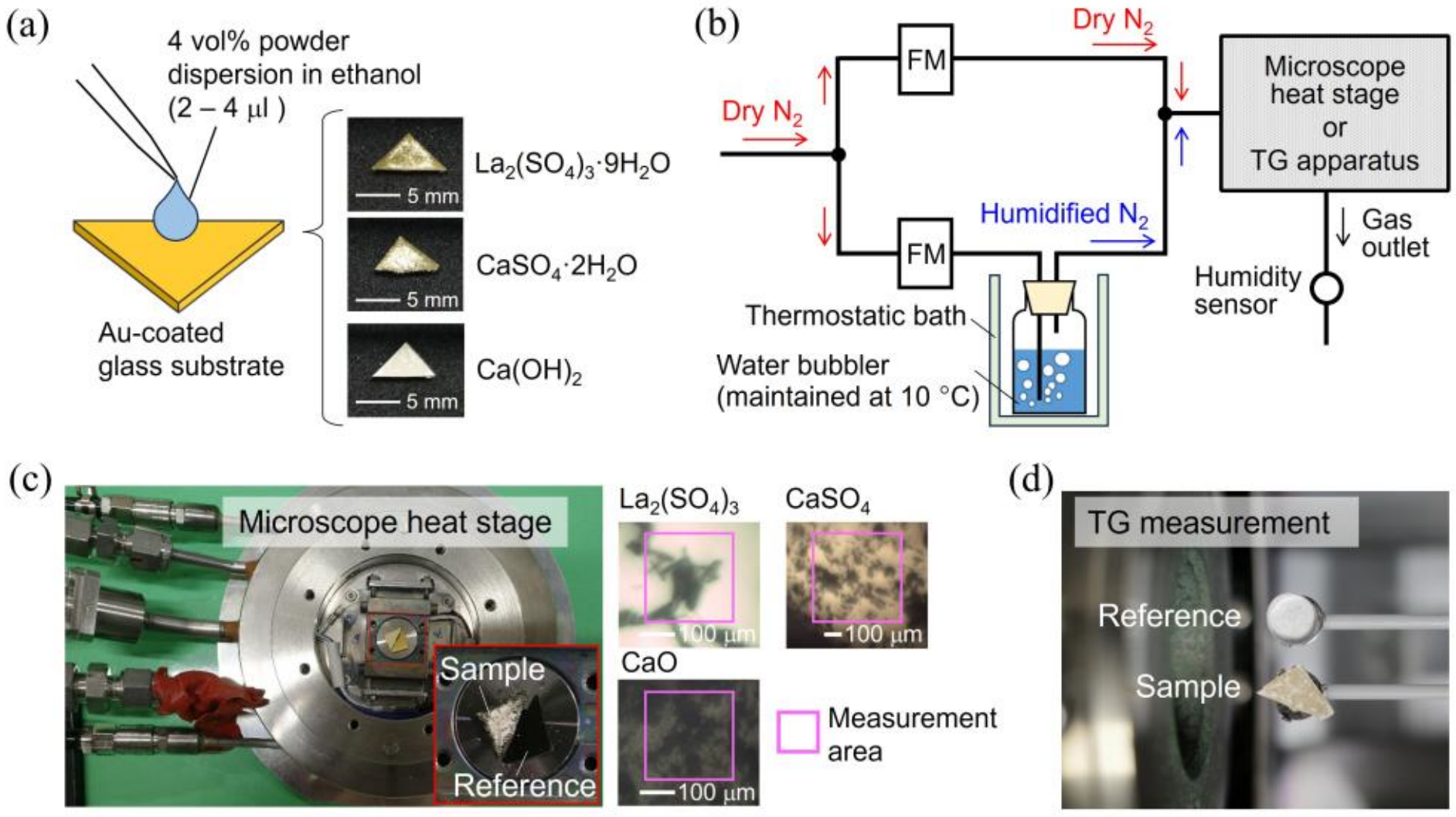

**Figure 1** Experimental workflow for in situ micro-FTIR measurements and corresponding TG measurements. (a) Preparation of powder samples on an Au-coated glass substrate and representative photographs of samples prepared using the same procedure. (b) Gas-flow system for supplying a mixture of dry and water-bubbled $N_2$ to either the microscope heat stage (S-100R) or the TG apparatus. The water bubbler was maintained at 10 °C, and the humidity of the outlet gas was monitored using a humidity sensor. FM denotes a needle-valve flow meter. (c) Microscope heat stage used for in situ micro-FTIR measurements, showing the sample and reference positions, together with optical micrographs indicating the measurement area for each material. (d) Sample and reference configuration used for TG measurements under corresponding hydration conditions.

## 2.2 In situ micro-FTIR measurements

Infrared spectra were acquired using a microscopic Fourier transform infrared spectrometer (FT/IR-6300 equipped with an IRT-5000 microscope, JASCO, Japan) combined with a gas-controllable microscope heat stage (S-100R, Pike Technologies, USA, Fig. 1(c)) fitted with $BaF_2$ windows. Spectra were acquired in transflection mode from powder layers deposited on Au-coated glass substrates, with a sample-free Au-coated glass substrate used as the reference. The external optical path between the microscope objective and the chamber window was exposed to ambient air; consequently, residual gas-phase $CO_2$ features remained near 2350 cm$^{-1}$. Spectra were collected with a spectral resolution of 4.0 cm$^{-1}$. For $La_2(SO_4)_3$ and $CaSO_4$, 15 scans were averaged to obtain each spectrum, whereas 30 scans were averaged for CaO. The resulting intervals between consecutive spectrum acquisitions were approximately 14 s for $La_2(SO_4)_3$ and $CaSO_4$ and 24 s for CaO.

Each deposited precursor layer was thermally dehydrated inside the environmental chamber under flowing dry $N_2$. The dry $N_2$ flow rates during dehydration reaction were 820 mL min$^{-1}$ for $La_2(SO_4)_3·9H_2O$ and $CaSO_4·2H_2O$ and 800 mL min$^{-1}$ for $Ca(OH)_2$. The deposited powder layers were heated to 300, 140, and 500 °C, respectively, to obtain $La_2(SO_4)_3$, $CaSO_4$, and CaO. These material-specific dehydration temperatures were selected to ensure complete dehydration based on the TG hydration–dehydration profiles shown in Supplementary Fig. S1. Following dehydration, each sample was cooled to and maintained at approximately 30 °C under the same dry $N_2$ flows.

Humidified $N_2$ was generated by passing $N_2$ through a water bubbler maintained at 10 °C. Hydration was initiated by switching the gas flow from dry $N_2$ to a combined flow of dry and humidified $N_2$. Hereafter, the corresponding flow rates are given in the order dry/humidified $N_2$. Flow rates of 800/20 mL min$^{-1}$ were used for hydration of $La_2(SO_4)_3$ and $CaSO_4$ whereas 650/150 mL min$^{-1}$ was used for hydration of CaO. Additional micro-FTIR and TG measurements were conducted at 700/100 mL min$^{-1}$ for $La_2(SO_4)_3$ and CaO. The $p_{H_2O}$ at the gas outlet was monitored continuously using a humidity sensor (SHT85, Sensirion, Switzerland). The measured outlet $p_{H_2O}$ values differed from the nominal values calculated assuming water vapor saturation at 10 °C and ideal mixing of dry and humidified $N_2$ streams. This discrepancy likely stems from residual moisture in the delivery system, non-ideal mixing, or minor ambient air leakage around the sensor assembly. Consequently, the measured values were used as operational indicators to monitor and compare atmospheres during measurements, rather than as exact local $p_{H_2O}$ values at the sample position.

## 2.3 Thermogravimetric measurements

Thermogravimetric (TG) measurements were performed using a Thermo plus EVO II TG8120 (Rigaku, Japan). In conventional TG measurements, powder samples are typically placed in an aluminum pan. However, preliminary experiments showed that the apparent hydration kinetics depended on the amount of powder loaded into the pan (Supplementary Fig. S2). To minimize differences in sample geometry between the TG measurements and micro-FTIR measurements, the TG measurements were performed using powder layers deposited on Au-coated glass substrates (Fig. 1(d)) in the same manner as the FTIR samples.

The deposited powder layers were thermally dehydrated inside the TG apparatus using the same material-specific dehydration temperatures and dry $N_2$ flow rates as in the corresponding micro-FTIR measurements. Following dehydration, each sample was cooled to 30 °C and maintained under dry $N_2$. Hydration was initiated using the same dry/humidified $N_2$ flow rates as those used for the corresponding micro-FTIR measurements. The relative humidity and temperature at the gas outlet were monitored using the same humidity sensor as in the micro-FTIR measurements. To minimize baseline fluctuations associated with switching the gas flow, an aluminum pan containing $Al_2O_3$ powder was placed on the reference side of the TG balance. The aluminum pan was 5 mm in diameter and 2.5 mm in height, and the total mass of the reference assembly was adjusted to match that of the sample assembly consisting of the Au-coated glass substrate and deposited powder layer.

To quantify hydration progress, the mass increase measured by TG was converted into a hydration number, defined here as the number of water molecules taken up per formula unit of the dehydrated material. This conversion requires the amount of dehydrated material on the substrate. Because the small mass of the deposited powder could not be measured directly, it was determined indirectly from the measured mass loss and the known stoichiometry of the following complete dehydration reactions:

$$La_2(SO_4)_3{\cdot}9H_2O(s) \rightarrow La_2(SO_4)_3(s) + 9H_2O(g) \quad (4)$$

$$CaSO_4{\cdot}2H_2O(s) \rightarrow CaSO_4(s) + 2H_2O(g) \quad (5)$$

$$Ca(OH)_2(s) \rightarrow CaO(s) + H_2O(g) \quad (6)$$

Because the dehydration mass losses of the deposited powder layers were on the order of tens of micrograms, baseline drift could not be neglected. To separate the mass loss due to dehydration from this drift, baseline correction was applied to the TG curves. Linear baselines were fitted to the approximately linear regions before and after dehydration. The time at which the TG curve reached the midpoint between the two baselines was defined as $t_{TG50}$, and the vertical difference between the baselines at $t_{TG50}$ was taken as the dehydration mass loss, $\Delta m_{dehyd}$. Assuming complete dehydration of each starting material, the TG-derived hydration number, $x(t)$, was calculated as

$$x(t) = n\,\frac{\Delta m_{\mathrm{hyd}}(t)}{\Delta m_{\mathrm{dehyd}}} \quad (7)$$

where $\Delta m_{hyd}(t)$ is the mass increase from the dehydrated state at reaction time $t$, $\Delta m_{dehyd}$ is the mass loss measured during dehydration, and $n$ is the number of water molecules released in the corresponding dehydration reaction shown in Eqs. (4)–(6). For CaO, the TG-derived hydration number is an apparent value because the measured mass increase may include contributions from both $H_2O$ uptake during hydration and $CO_2$ uptake during carbonation, as explained in Section 3.2. Thus, it therefore does not represent the progress of hydration alone.

## 3. Results

### 3.1 Dehydration behavior of powder layers deposited on substrates

Figure 2 shows the TG curves obtained during heating of $La_2(SO_4)_3{\cdot}9H_2O$, $CaSO_4{\cdot}2H_2O$, and $Ca(OH)_2$ deposited on the substrates. All three materials underwent dehydration reactions shown in Eqs. (4)–(6). The observed dehydration temperature ranges were similar to those obtained from conventional powder samples placed directly in aluminum pans (Supplementary Fig. S1). Thus, despite the small amounts of deposited powder, the TG measurements captured the expected dehydration behavior. $La_2(SO_4)_3{\cdot}9H_2O$ showed a broad multistep mass decrease between 100–240 °C, consistent with the reported formation of an amorphous intermediate during dehydration.[26]

As illustrated in Fig. 2, the dehydration mass loss was determined from the vertical difference between the pre- and post-dehydration baselines at $t_{TG50}$. The resulting dehydration mass losses, $\Delta m_{dehyd}$, were 27.2 μg for $La_2(SO_4)_3{\cdot}9H_2O$, 19.8 μg for $CaSO_4{\cdot}2H_2O$, and 41.3 μg for $Ca(OH)_2$. These values were used as internal references to convert the mass changes during the hydration measurements, $\Delta m_{hyd}(t)$, into the TG-derived hydration number, $x(t)$, using Eq. (7).

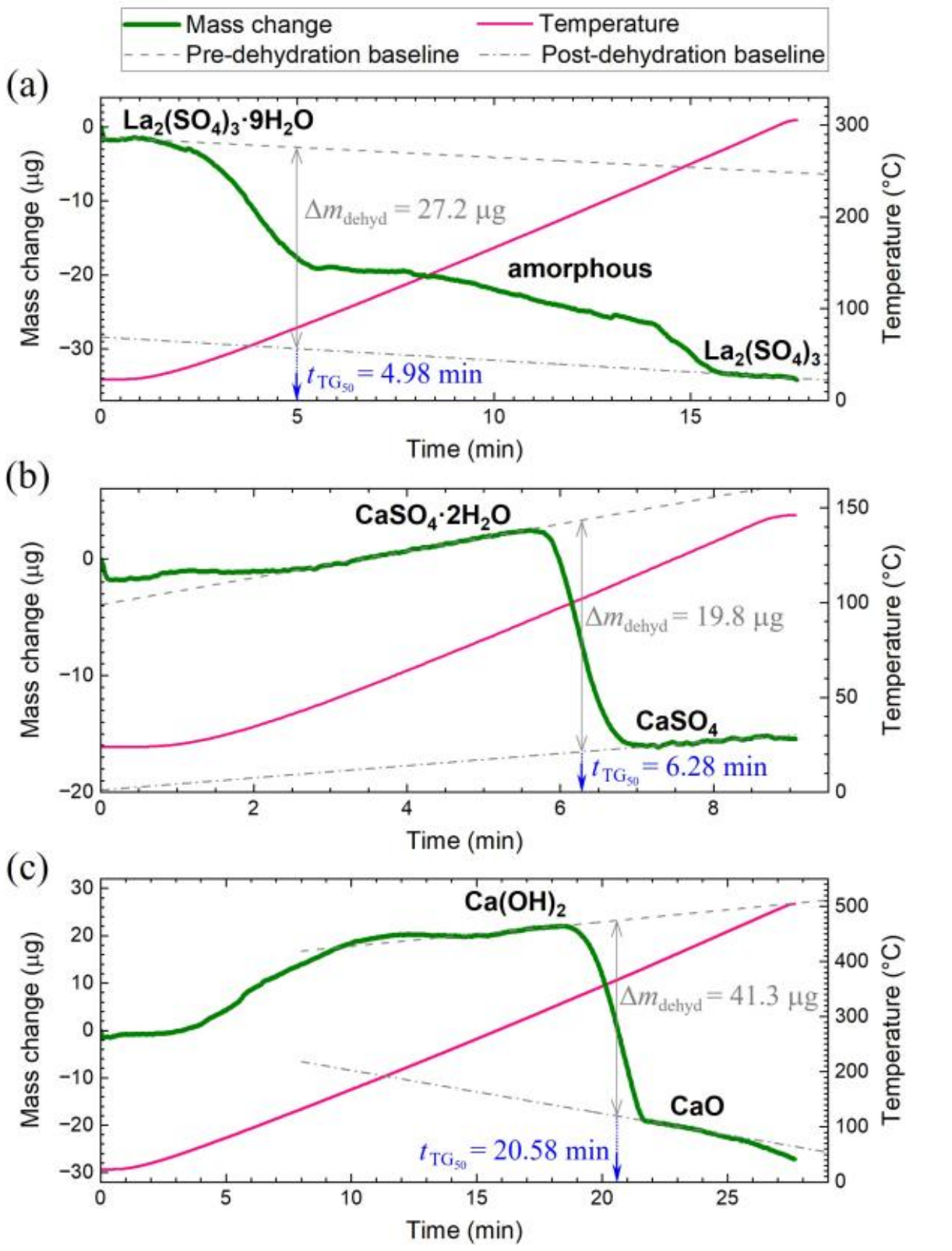

**Figure 2** TG curves obtained during dehydration of powder layers deposited on substrates: (a) $La_2(SO_4)_3{\cdot}9H_2O$, (b) $CaSO_4{\cdot}2H_2O$, and (c) $Ca(OH)_2$. The gray dashed and dash-dotted lines represent the pre- and post-dehydration baselines, respectively. The blue downward arrows indicate the $t_{TG50}$, defined as the time at which the TG curve reaches the midpoint between the two baselines. The gray double-head arrows indicate $\Delta m_{dehyd}$, the dehydration mass loss for reactions (4)–(6), defined as the vertical differences between the two baselines at $t_{TG50}$.

## 3.2 In situ micro-FTIR observation of hydration reactions

Figure 3 shows representative time-resolved FTIR spectra acquired during the hydration reactions of $La_2(SO_4)_3$, $CaSO_4$, and CaO shown in Eqs. (1)–(3). The spectra labeled "Dry $N_2$" were acquired after the dehydration treatment and subsequent cooling to 30 °C under dry $N_2$, immediately before switching to humidified $N_2$.

FTIR spectra of the starting materials, $La_2(SO_4)_3{\cdot}9H_2O$, $CaSO_4{\cdot}2H_2O$, and $Ca(OH)_2$, and their dehydrated products immediately after heating are provided in Supplementary Fig. S3. The water- and hydroxyl-related absorption bands markedly decreased after heating, confirming the dehydration of these starting materials. During subsequent cooling under dry $N_2$, these bands partially recovered, indicating the partial rehydration had occurred before humidification. Although the samples were not fully dehydrated at the start of humidification, the spectra obtained under dry $N_2$ at 30 °C immediately before humidification were used as the reference state for analyzing the subsequent changes.

For $La_2(SO_4)_3$ and $CaSO_4$, the broad absorption bands at approximately 3000–3700 $cm^{-1}$ and the bands near 1630 $cm^{-1}$ ($La_2(SO_4)_3$) and 1615 $cm^{-1}$ ($CaSO_4$) were assigned to the O–H stretching and H–O–H bending vibrations of molecular water, respectively. The intense bands below approximately 1300 $cm^{-1}$ were attributed to internal vibrations of $SO_4^{2-}$, while their overtone and combination bands were observed in the 1900–2300 $cm^{-1}$ region.[28,29] During the hydration reaction, the water-related bands progressively increased, whereas the sulfate-related bands retained their overall spectral profiles. These changes demonstrate that the hydration progress can be traced from the increasing absorption bands of molecular water. The hydration steps from $La_2(SO_4)_3$ to $La_2(SO_4)_3{\cdot}H_2O$ and from $CaSO_4$ to $CaSO_4{\cdot}0.5H_2O$ have been reported to involve the insertion of molecular water into structural channels while largely retaining the respective cation–sulfate frameworks.[30,31] The observed increases in both the O–H stretching and H–O–H bending bands are consistent with these hydration mechanisms involving molecular-water insertion.

In contrast, the hydration reaction of CaO produced a sharp band near 3640 $cm^{-1}$, which was assigned to the O–H stretching vibration of hydroxyl groups in $Ca(OH)_2$. Unlike $La_2(SO_4)_3$ and $CaSO_4$, no distinct H–O–H bending band was observed, indicating hydroxide formation rather than molecular-water insertion. In addition, carbonate-related bands near 1450 and 875 $cm^{-1}$, assigned to $CO_3^{2-}$ asymmetric stretching and out-of-plane bending vibrations, respectively, increased during the measurement.[32] These changes indicate partial carbonation, represented by Eq. (8), presumably caused by reaction with trace $CO_2$ in the atmosphere.[4]

$$CaO(s) + CO_2\,(g) \rightarrow CaCO_3(s) \tag{8}$$

Thus, the in situ time-resolved FTIR spectra distinguished molecular-water insertion in the sulfate materials from hydroxide formation and concurrent partial carbonation in CaO.

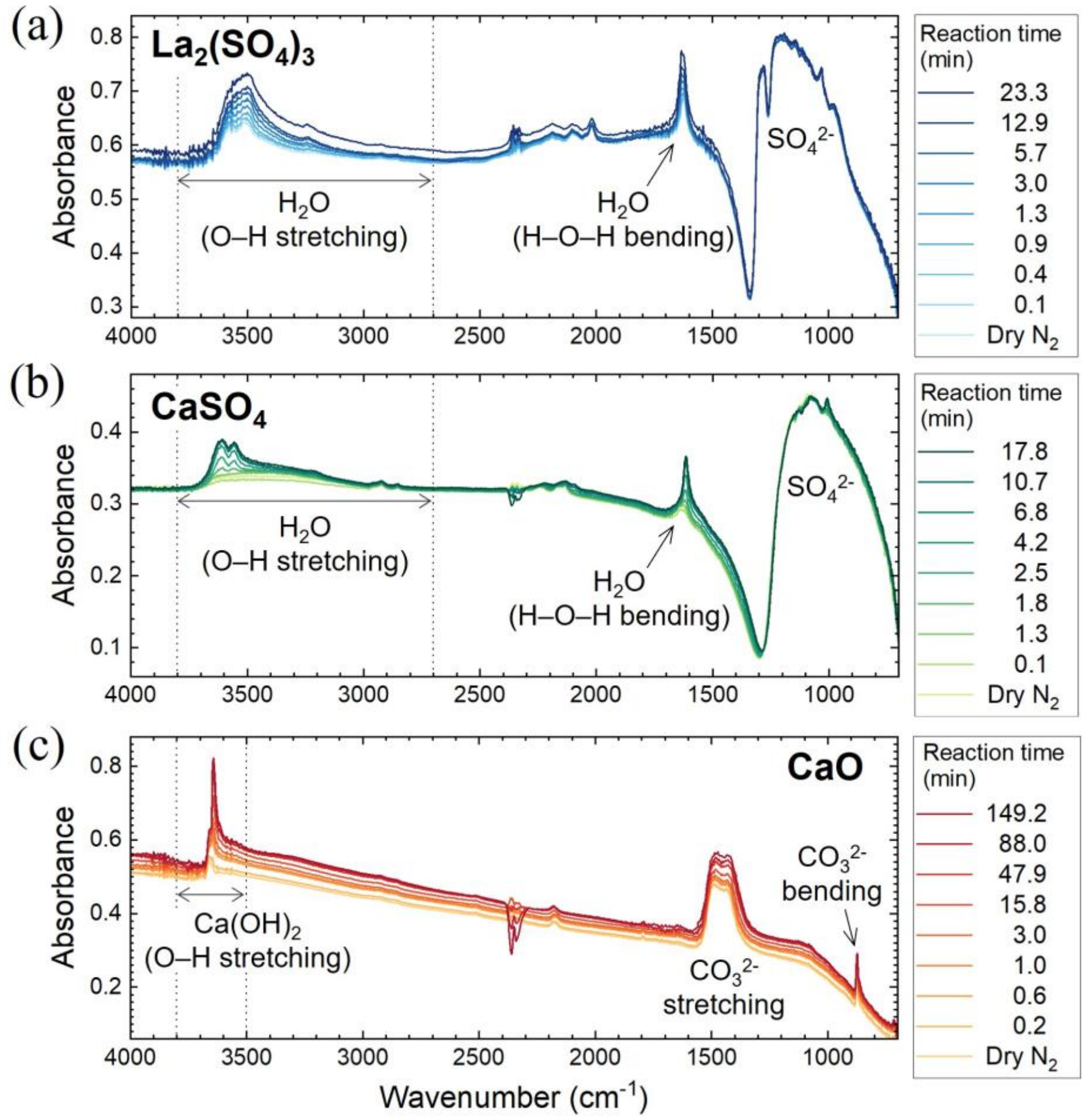

**Figure 3** Representative in situ FTIR spectra acquired at 30 °C during the hydration reactions under humidified $N_2$: (a) $La_2(SO_4)_3$, (b) $CaSO_4$, and (c) CaO. The spectra labeled "Dry $N_2$" were acquired immediately before humidification and were used as the reference spectra for analyzing the subsequent changes. Reaction time was defined as the elapsed time after switching the gas flow from dry $N_2$ to humidified $N_2$. The sharp features near 2350 $cm^{-1}$ originate from atmospheric $CO_2$ in the optical path and are not sample-related. The double-headed arrows indicate the wavenumber ranges used for quantification as band areas: 2700–3800 $cm^{-1}$ for $La_2(SO_4)_3$ and $CaSO_4$, and 3500–3800 $cm^{-1}$ for CaO. The integrated areas represent the positive absorbance above a straight-line baseline connecting the absorbance values at the lower and upper wavenumber limits of each integration range.

## 3.3 Comparison of hydration kinetics measured by micro-FTIR and TG

The time-dependent changes in the water- and hydroxyl-related absorption bands (Fig. 3) were quantified by integrating their absorbance to evaluate hydration progress. Absorption band areas were determined after subtracting a two-point linear baseline. For each integration range, a straight-line baseline was drawn between the absorbance values at its lower and upper wavenumber limits. The positive absorbance above this baseline was integrated over 2700–3800 cm$^{-1}$ for $La_2(SO_4)_3$ and $CaSO_4$ and over 3500–3800 cm$^{-1}$ for CaO.

Figure 4 compares the time-dependent IR absorption-band areas with the corresponding TG-derived hydration numbers. The upper panels show the water-vapor-pressure $p_{H_2O}$ profiles measured at the gas outlet during the micro-FTIR and TG experiments. Although slight differences were observed between the $p_{H_2O}$ profiles because of differences in the experimental apparatus, $p_{H_2O}$ stabilized shortly after gas switching and remained nearly constant throughout each hydration measurement. The lower panels compare the water- and hydroxyl-related O–H stretching band areas for $La_2(SO_4)_3$, $CaSO_4$, and CaO with the corresponding TG-derived hydration numbers.

Because micro-FTIR does not directly provide an absolute hydration number, the vertical-axis scales were adjusted for visual comparison so that the IR absorption-band-area increase from $t = 0$ to $t_{plateau}$, the practical reaction-completion time, matched the TG-derived hydration number at $t_{plateau}$. The $t_{plateau}$, was estimated from the smoothed TG-derived hydration number curve using a moving-average-based plateau detection algorithm. The plateau-detection method is explained in Fig. S4. This adjustment affected only the graphical axis scales; the original signal values were retained, and the FTIR signals were not converted into absolute hydration numbers.

Although not shown in Fig. 4, the integrated H–O–H bending band areas for both $La_2(SO_4)_3$ and $CaSO_4$ exhibited temporal evolutions similar to those of the corresponding O–H stretching band areas. The O–H stretching band areas were therefore used as representative water-related FTIR signals for comparison with the TG-derived hydration numbers.

For $La_2(SO_4)_3$, the water-related absorption band area increased gradually over several tens of minutes, closely following the increase in the TG-derived hydration number. $CaSO_4$ showed a rapid initial increase during the first few minutes, followed by an approach toward nearly constant values over the next few tens of minutes. In contrast, CaO hydration proceeded more gradually than that of the sulfate materials. Although the hydroxyl-related FTIR signal and TG-derived hydration number differed during the initial and intermediate stages, both showed a gradual increase over more than 200 min. Thus, despite some differences in the profile details, FTIR captured the overall hydration trend and prolonged timescale observed by TG for CaO.

Note that switching from dry to humidified $N_2$ produced a humidity-dependent background change, particularly in the 2700–3800 cm$^{-1}$ range (Supplementary Fig. S5). Thus, during the initial stage after gas switching, while $p_{H_2O}$ was rising, the absorption band areas of $La_2$ $(SO_4)_3$ and $CaSO_4$ may be overestimated because of additional absorption of gas-phase water vapor. However, at the $p_{H_2O}$ reached under the 800/20 mL min$^{-1}$ condition, the background absorption area in this range was estimated to be only 1–2 cm$^{-1}$ (Supplementary Fig. S5). This contribution was small relative to the changes in the sample signals and is unlikely to affect the interpretation of the hydration profiles.

Overall, straightforward integration of the water- and hydroxyl-related bands was sufficient to distinguish the different hydration timescales of the three materials. An additional measurement of $La_2(SO_4)_3$ was performed at a higher $p_{H_2O}$ and indicated a more rapid increase in both the water-related absorption band area and TG-derived hydration number (Supplementary Fig. S6). These results indicate that straightforward band area analysis can track relative differences in hydration kinetics arising from both material selection and reaction conditions.

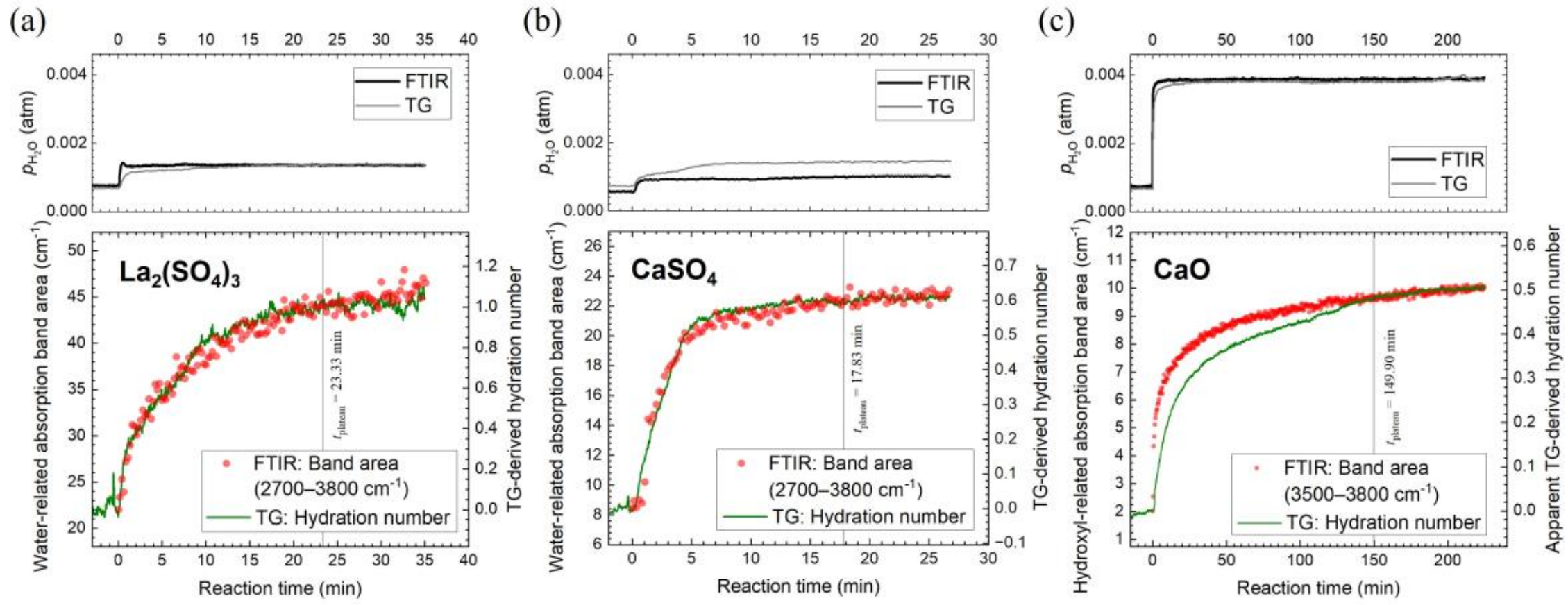

**Figure 4** Comparison of hydration kinetics measured by in situ micro-FTIR spectroscopy and TG analysis for (a) $La_2(SO_4)_3$, (b) $CaSO_4$, and (c) CaO. Dry/humidified $N_2$ flow rates were 800/20 mL $min^{-1}$ for $La_2(SO_4)_3$ and $CaSO_4$ and 650/150 mL $min^{-1}$ for CaO. The upper panels compare the water vapor pressure, $p_{H_2O}$, measured during the FTIR and TG experiments. For CaO, the TG-derived hydration number is an apparent value because the measured mass increase may include contributions from both hydration and carbonation. The FTIR measurements were conducted at 30.0 °C, whereas the temperature during the TG measurements was controlled near 30 °C. The mean temperatures during hydration were 29.5 °C for $La_2(SO_4)_3$, 29.4 °C for $CaSO_4$, and 29.7 °C for CaO. The lower panels compare the water-related absorption band areas for $La_2(SO_4)_3$ and $CaSO_4$ and the hydroxyl-related absorption band area for CaO with the corresponding TG-derived hydration numbers. The absorption band areas were determined after subtraction of a linear baseline connecting the endpoints of each integration range: 2700–3800 $cm^{-1}$ for $La_2(SO_4)_3$ and $CaSO_4$ and 3500–3800 $cm^{-1}$ for CaO. A reaction time of 0 min corresponds to switching the gas flow from dry $N_2$ to the prescribed dry/humidified $N_2$ flow rate combination. For visual comparison, the vertical-axis scales were adjusted so that the increase in the absorption-band area from $t$ = 0 to $t_{plateau}$ had the same vertical span as the TG-derived hydration-number increase to $t_{plateau}$. The TG-derived hydration number at $t_{plateau}$ was calculated as the mean within a 5-min window centered on $t_{plateau}$. The $t_{plateau}$ values used for this adjustment are indicated in the panels. For each material, the profiles are shown from $t$=0 to the practical TG endpoint, $t_{end} = 1.5t_{plateau}$.

## 3.4 Sensitivity of CaO hydration behavior to humidification conditions

Because CaO showed a relatively large difference between the FTIR and TG profiles in Fig. 4(c), additional measurements were conducted to examine the sensitivity of its reaction behavior to the humidification conditions. Figure 5 compares the CaO reaction behavior measured by TG and in situ micro-FTIR under dry/humidified $N_2$ flow rates of 650/150 and 700/100 mL min$^{-1}$. The 700/100 mL min$^{-1}$ data are shown in Supplementary Fig. S7. The 650/150 mL min$^{-1}$ data are the same as those presented in Fig. 4(c). Under both conditions, the $p_{H_2O}$ profiles increased rapidly after $t$=0 and then remained nearly constant (Fig. 5(a)). The stabilized $p_{H_2O}$ decreased from approximately 3.8×10$^{-3}$ atm under the 650/150 mL min$^{-1}$ condition to 3.1×10$^{-3}$ atm under the 700/100 mL min$^{-1}$ condition.

Despite this relatively small difference in $p_{H_2O}$, the apparent TG-derived hydration number showed markedly different temporal profiles under the two conditions (Fig. 5(b)). Under the 650/150 mL min$^{-1}$ condition, it reached approximately 0.38 at 60 min, whereas under the 700/100 mL min$^{-1}$ condition, it increased more slowly, reaching approximately 0.20 over the same period.

The FTIR measurements showed increases in both the hydroxyl- and carbonate-related absorption-band areas (Fig. 5(c)). The band-area changes are shown relative to their respective values at $t$=0, as $\Delta A(t) = A(t) - A(0)$. The carbonate-related signal exhibited distinctly different temporal profiles under the two conditions. It increased sharply during the initial stage under the 700/100 mL min$^{-1}$ condition, whereas it increased gradually throughout the measurement period under the 650/150 mL min$^{-1}$ condition. Thus, the condition with the more rapid initial increase in the carbonate-related signal showed the smaller increase in the TG-derived hydration number.

Under the 700/100 mL min$^{-1}$ condition, the carbonate-related FTIR signal increased more rapidly, whereas the apparent TG-derived hydration number remained lower than under the 650/150 mL min$^{-1}$ condition. If carbonation only added mass to the TG signal, the apparent hydration number would also be expected to increase as carbonation progressed. However, it was lower under the condition with the greater carbonate-related FTIR signal. This suggests that carbonation may also inhibit hydration. This interpretation is consistent with previous reports that a $CaCO_{33}$ layer formed on CaO can impede subsequent hydration and reduce the hydration capacity.[33,34] If such a layer formed during the early stage of the reaction, it may have restricted $H_2O$ access to the underlying CaO, suppressing further hydration and limiting the subsequent mass gain.

The initial $p_{H_2O}$ profiles during gas-replacement differed between the FTIR and TG measurements, owing to differences in the gas volumes of the two setups. Taking into consideration the sensitivity of CaO reaction behavior to the gas-phase atmosphere, these differences may have affected initial carbonate formation and contributed to the deviation between the FTIR and TG profiles in Fig. 4(c). Surface-carbonation-induced inhibition may also have contributed to the TG-derived hydration number remaining below 1 (Fig. 4(c)), the value expected for complete conversion to $Ca(OH)_2$. However, the formation of a surface carbonate layer and its inhibiting effect were not directly confirmed in this study.

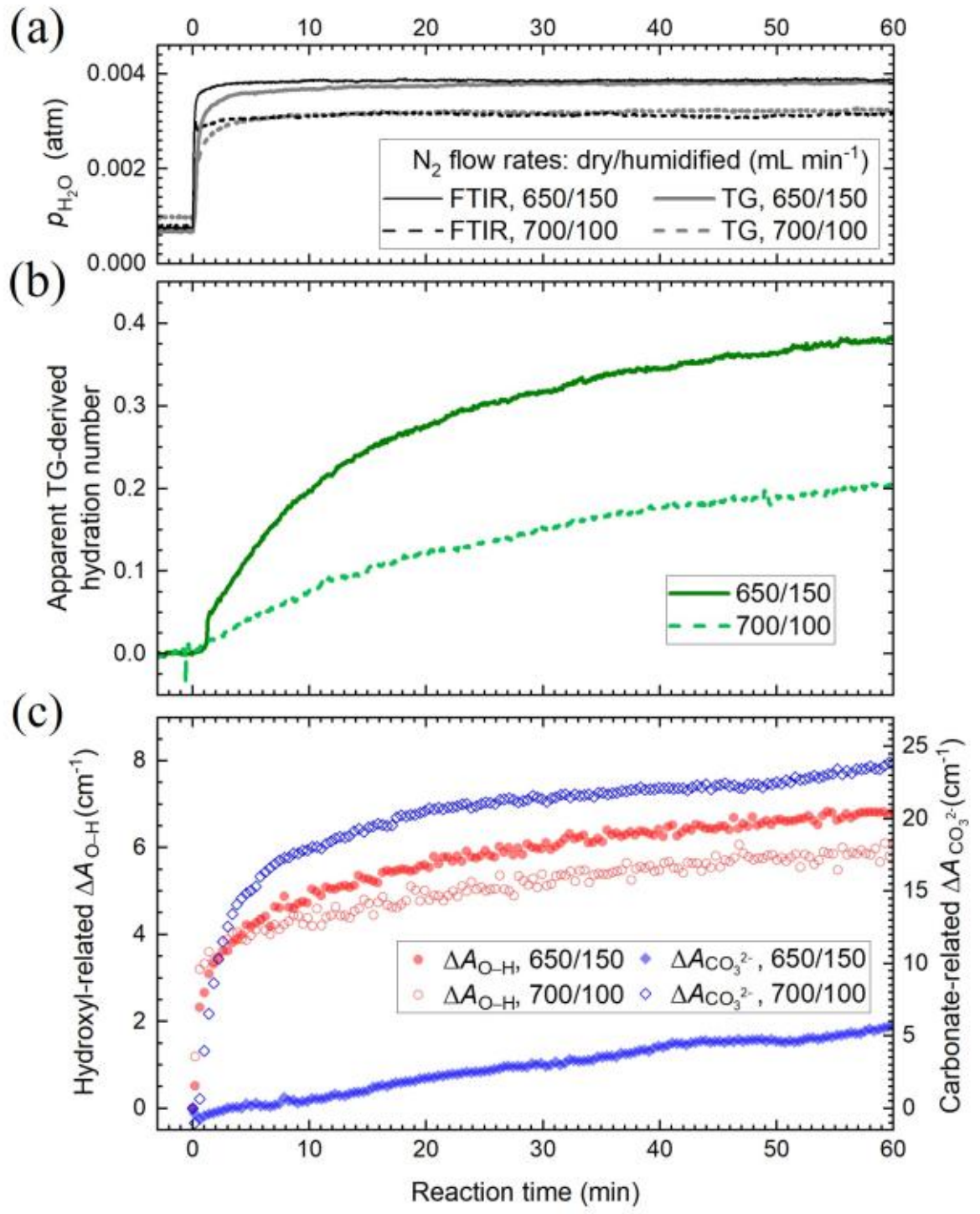

**Figure 5** CaO hydration behavior under different dry/humidified $N_2$ flow rate conditions. The dry/humidified $N_2$ flow rates were 650/150 and 700/100 mL min$^{-1}$. A reaction time of 0 min ($t = 0$) corresponds to switching the gas flow from dry $N_2$ to the mixture of dry and humidified $N_2$. (a) Water vapor pressure, $p_{H_2O}$, measured at the gas outlet during the FTIR and TG experiments. (b) Apparent TG-derived hydration numbers. Solid and dashed lines indicate the 650/150 and 700/100 mL min$^{-1}$ conditions, respectively. (c) Changes in the hydroxyl-related absorption band area integrated over 3500–3800 cm$^{-1}$ and the carbonate-related absorption band area integrated over 1200–1600 cm$^{-1}$. For each integration range, a straight-line baseline was drawn between the absorbance values at its lower and upper wavenumber limits, and the positive absorbance above this baseline was integrated. Changes in the band areas were calculated relative to the respective values at $t$=0, as $\Delta A(t) = A(t) - A(0)$. Filled and open symbols indicate the 650/150 and 700/100 mL min$^{-1}$ conditions, respectively. The 650/150 mL min$^{-1}$ datasets are the same as those presented in Fig. 4(c).

## 4. Discussion

Overall, straightforward integration of IR absorption-band areas closely followed the TG-derived hydration trends for $La_2(SO_4)_3$ and $CaSO_4$. For CaO, the hydroxyl-related FTIR signal and TG-derived hydration number differed more noticeably during the initial and intermediate stages, although both measurements captured the overall gradual increase over more than 200 min. This deviation may partly reflect the sensitivity of CaO to the atmosphere, together with differences in the initial $p_{H_2O}$ profiles during gas switching in the FTIR and TG setups (Fig. 5). Previous ATR-FTIR studies used spectral changes to follow the hydration of cementitious materials in contact with liquid water.[20–22] Consistent with previous ATR-FTIR studies showing that integral band areas can provide qualitative measures of hydration progress,[22] the present results demonstrate that straightforward band area analysis can also track water-vapor-driven gas–solid hydration across materials with different hydration mechanisms and timescales.

Under the conditions examined in this study, micro-FTIR successfully tracked hydration reactions occurring over timescales of several tens to several hundreds of minutes. The measurements were performed on powder layers comprising a few layers of particles, each on the order of tens of micrometers in size. The absorption bands used to monitor hydration could be distinguished without substantial overlap. However, the applicability of this approach to samples with different particle sizes or layer thicknesses, or with more strongly overlapping bands, has not yet been evaluated. In transflection measurements, the incident light passes through the powder layer, reflects from the substrate, and passes through the layer again. If the layer is too thick or the particles are too large, the reflected signal may be too weak for reliable measurement because less light reaches the substrate and returns to the detector. In addition, for transflection measurements, absorbance does not necessarily increase linearly with sample-layer thickness, as predicted by the Beer–Lambert law.[24] Therefore, the effective optical path length cannot be assumed to be simply twice the physical layer thickness, making it difficult to convert absorbance directly into an absolute water content or hydration number. The approach is therefore suited to tracking relative reaction progress, whereas determining absolute hydration amounts requires an independent quantitative reference. Further work is needed to establish the range of sample structures and reaction conditions over which the method can be applied reliably.

Despite these limitations, the microscopic, non-contact configuration of micro-FTIR enables reaction progress to be monitored within a sub-mm$^2$ area. This spatial resolution could facilitate mapping of reaction kinetics across powder layers. Moreover, automated point-by-point analysis of combinatorial samples arranged in an array on a substrate could provide a high-throughput experimental approach for evaluating the hydration kinetics of various candidate materials identified through computational screening. Because measurements can be made using only a small amount of sample, this approach may be particularly useful when materials are available in limited quantities. Finally, FTIR can distinguish absorption bands associated with different chemical species and track their evolution during reaction, potentially providing insight into reaction mechanisms, particularly when competing reactions occur simultaneously, as in the hydration and carbonation reactions observed in this study. Although demonstrated here for hydration, the approach may also be applicable to other gas–solid reactions when the reacting or product species exhibit distinguishable infrared bands.

## 5. Conclusion

This study shows that time-resolved transflection micro-FTIR with straightforward integration of diagnostic absorption bands can be used to evaluate the relative kinetics of gas–solid hydration in thin powder layers. For $La_2(SO_4)_3$, $CaSO_4$, and CaO, the FTIR band-area profiles captured the overall hydration behavior measured by TG, although some deviations were observed for CaO. Hydration reactions occurring over timescales of several tens to several hundreds of minutes were tracked under the conditions examined, using powder layers comprising a few layers of particles on the order of tens of micrometers in size.

FTIR distinguished molecular-water uptake in $La_2(SO_4)_3$ and $CaSO_4$ from hydroxide formation in CaO. For CaO, it also tracked carbonate-related signals separately from hydroxyl-related signals, revealing changes associated with concurrent carbonation. The CaO reaction behavior varied with the humidification conditions, and differences in the initial water-vapor-pressure profiles between the FTIR and TG setups may have contributed to the deviation between their profiles. These results show that FTIR can track competing reactions through signals associated with different chemical species.

The method evaluates relative reaction progress rather than absolute hydration amounts. Because the effective optical path length in transflection measurements is unknown, FTIR band areas cannot be directly converted into absolute water contents or hydration numbers. Absolute quantification therefore requires an independent quantitative reference. The applicability of the approach to samples with different particle sizes or layer thicknesses, and to materials with more strongly overlapping bands, remains to be established. Nevertheless, its microscopic, non-contact configuration allows measurements within sub-$mm^2$ areas using small amounts of sample. Automated mapping and point-by-point analysis of combinatorial sample arrays could enable high-throughput kinetic screening of computationally identified candidate materials. Tracking the evolution of individual chemical species may also provide insight into reaction mechanisms, particularly when competing reactions occur.

## 6. Acknowledgement

This work was supported by a JSPS Grant-in-Aid for Young Scientists (26K17642), The Tatematsu Research Grant, The Naito Research Grant, and The NAGAI Foundation of Science & Technology. This work was the result of using research equipment of the Equipment Sharing Division, Organization for Co-Creation Research and Social Contributions, Nagoya Institute of Technology, shared under the JST EPOCH program (Grant Number JPMJEP2610). The authors acknowledge the use of ChatGPT (OpenAI) for English language editing and improvements to manuscript readability. The authors reviewed and take full responsibility for the final content.

# Supporting Information

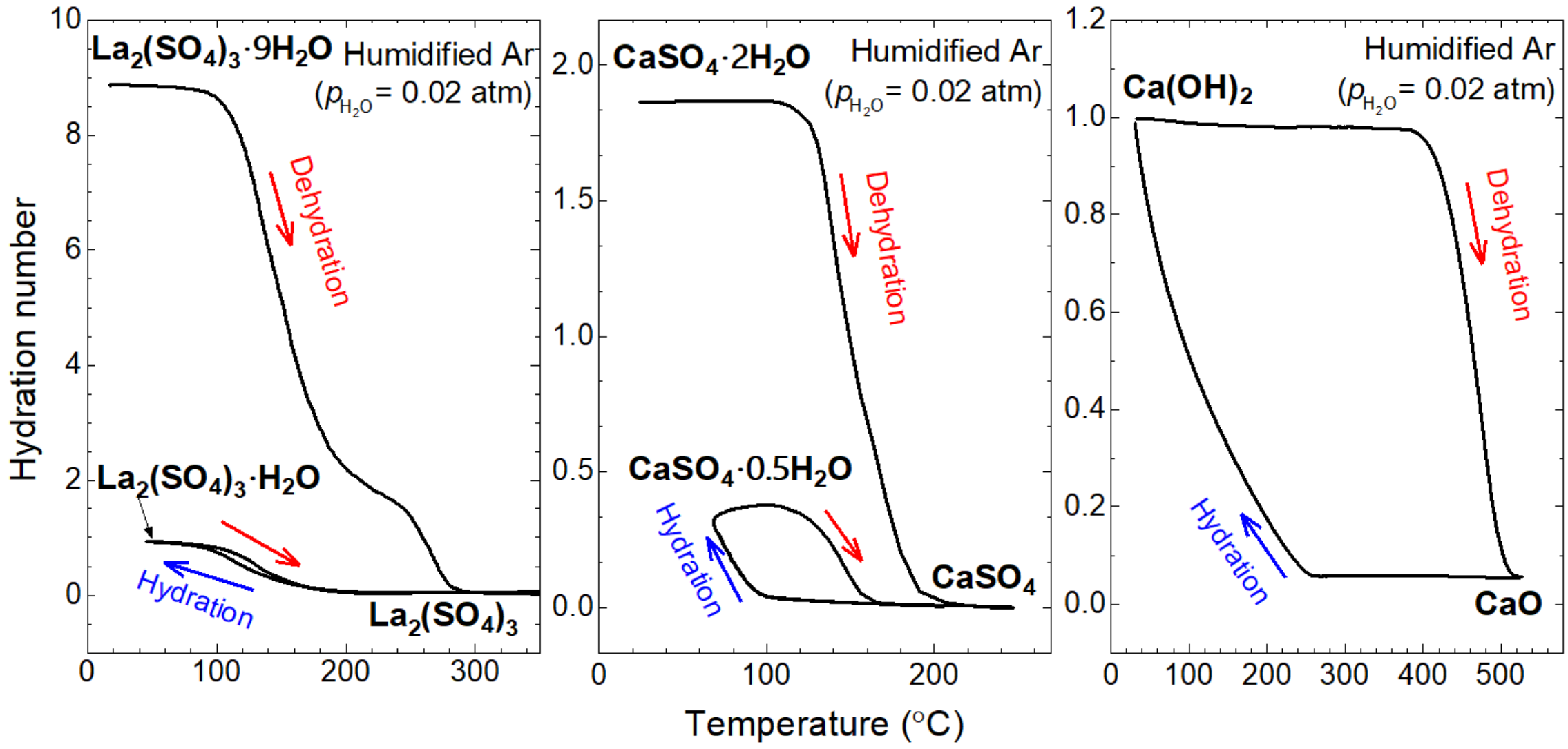

**Figure S1** Representative TG curves of conventional powder samples placed directly in aluminum pans under humidified Ar ($p_{H_2O}$= 0.02 atm): (a) $La_2(SO_4)_3$, (b) $CaSO_4$, and (c) CaO systems. The initial dehydration curves show the conversion of $La_2(SO_4)_3 \cdot 9H_2O$, $CaSO_4 \cdot 2H_2O$, and $Ca(OH)_2$ into $La_2(SO_4)_3$, $CaSO_4$, and CaO, respectively, as shown in Eqs. (4)–(6). The subsequent curves show the reversible reactions between $La_2(SO_4)_3$ and $La_2(SO_4)_3 \cdot H_2O$, between $CaSO_4$ and $CaSO_4 \cdot 0.5H_2O$, and between CaO and $Ca(OH)_2$, as shown in Eqs. (1)–(3). The red and blue arrows indicate the dehydration and hydration directions, respectively. Hydration numbers were calculated from the TG mass changes and the initial sample masses based on the corresponding reaction stoichiometries.

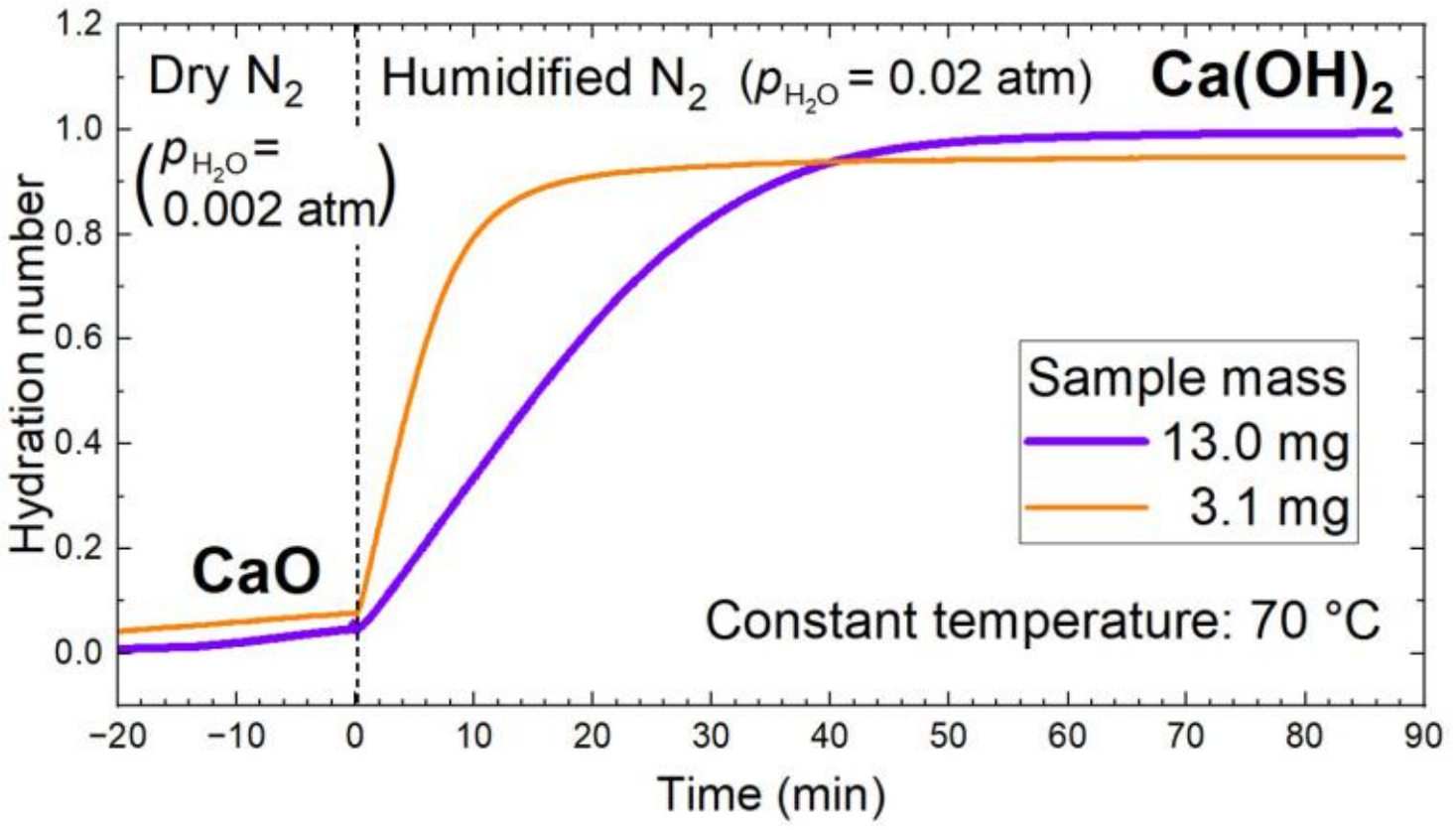

**Figure S2** Effect of sample mass on the apparent hydration kinetics measured by TG analysis. Hydration curves of CaO were obtained using powder samples with initial masses of 13.0 and 3.1 mg placed in an aluminum pan, 5 mm in diameter and 2.5 mm in height. Hydration reaction was initiated at 70 °C by switching the gas flow from dry $N_2$ to humidified $N_2$ ($p_{H_2O}$ = 0.02 atm). Although CaO exhibited the slowest hydration reaction among the three materials investigated in this study, its apparent hydration kinetics depended on the packed sample mass. This observation led to the use of powder layers deposited on substrates in the TG measurements, allowing comparison with the hydration kinetics evaluated by in situ micro-FTIR.

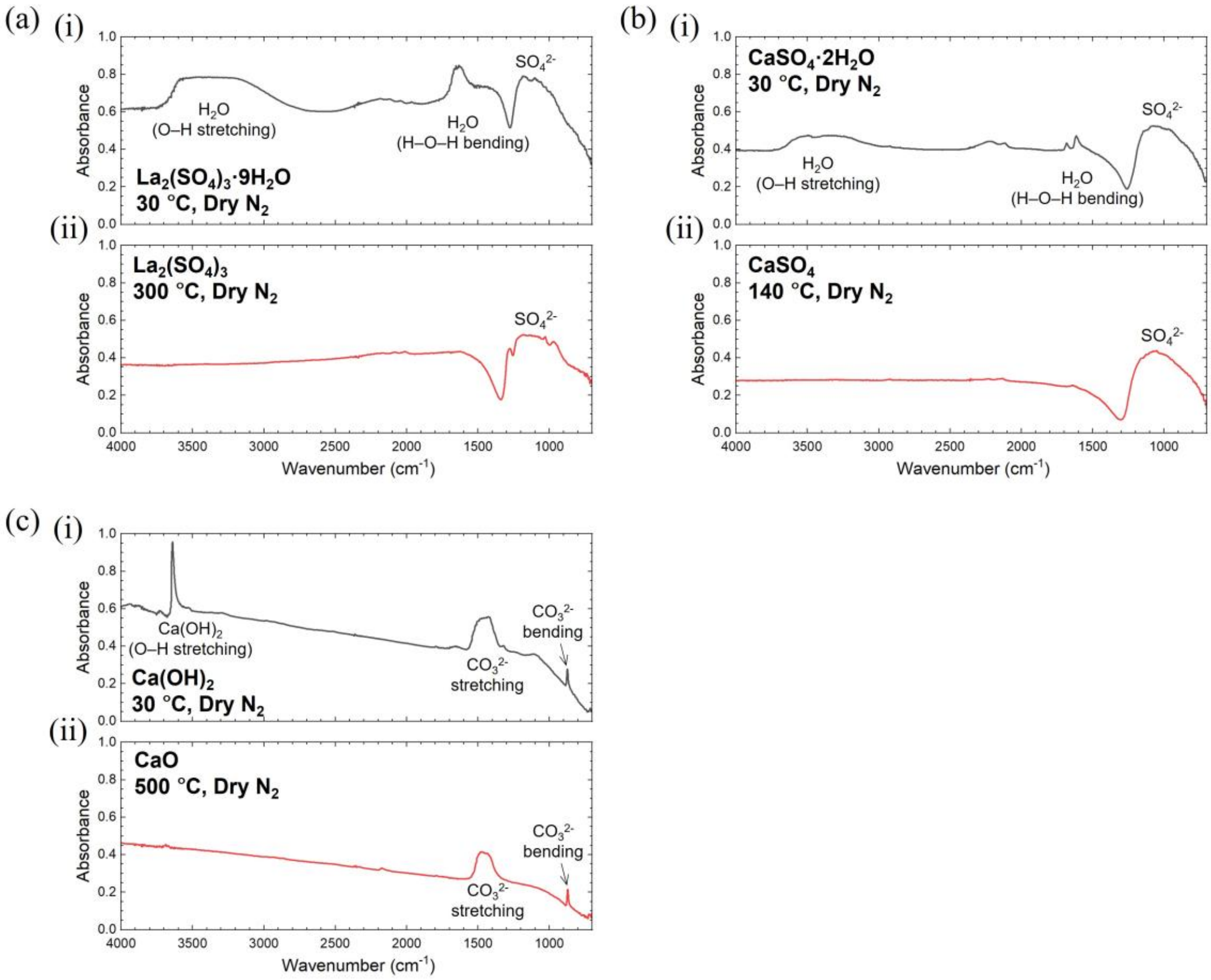

**Figure S3** Representative FTIR spectra of (i) the starting materials, $La_2(SO_4)_3 \cdot 9H_2O$, $CaSO_4 \cdot 2H_2O$, and $Ca(OH)_2$, and (ii) the corresponding dehydrated phases, $La_2(SO_4)_3$, $CaSO_4$, and CaO, acquired before the in situ hydration measurements. The O–H stretching and H–O–H bending bands of molecular water were markedly reduced by dehydration of $La_2(SO_4)_3 \cdot 9H_2O$ and $CaSO_4 \cdot 2H_2O$, whereas $Ca(OH)_2$ was characterized by a sharp O–H stretching band of hydroxyl groups.

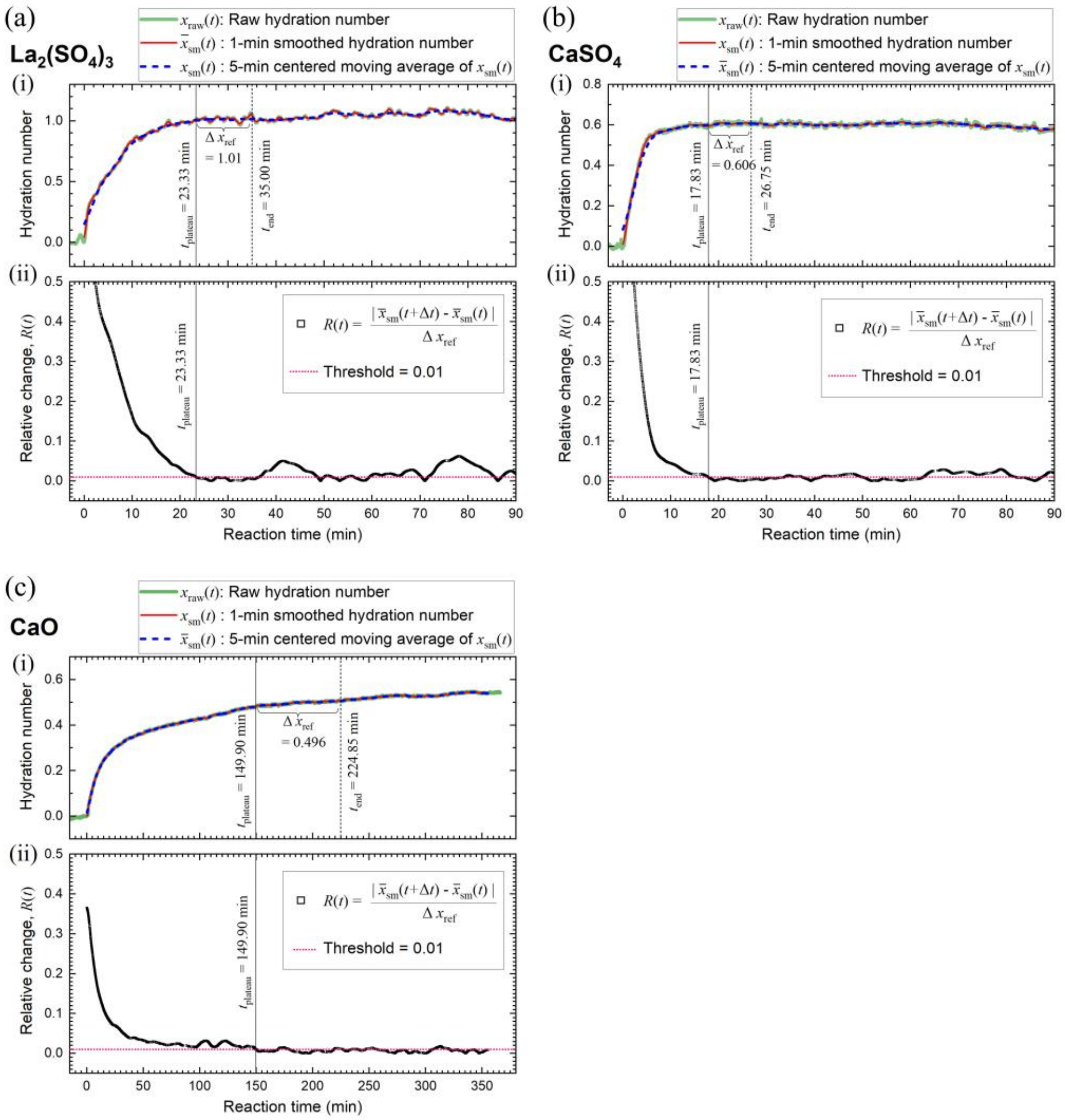

**Figure S4** Moving-average-based plateau detection and determination of the reference hydration-number change for (a) $La_2(SO_4)_3$, (b) $CaSO_4$, and (c) CaO. In panel (i), the raw TG-derived hydration number, $x_{raw}(t)$, is shown with the 1-min-smoothed hydration number, $x_{sm}(t)$, and its 5-min centered moving average, $\bar{x}_{sm}(t)$. In panel (ii), the relative change, $R(t)$, is shown as a function of reaction time:

$$R(t) = \frac{|\bar{x}_{sm}(t+\Delta t) - \bar{x}_{sm}(t)|}{\Delta x_{ref}}$$

where $\Delta t$ =10 min and $\Delta x_{ref}$ is the mean of $\bar{x}_{sm}(t)$ over the post-plateau interval from $t_{plateau}$ to $t_{end}$. The baseline was set such that $x(t_0) = 0$. The practical reaction-completion time, $t_{plateau}$, was defined as the first time point at which $R(t)<0.01$. The endpoint was set to $t_{end} = 1.5\ t_{plateau}$, using the nearest measured time point. Because $\Delta x_{ref}$ depends on $t_{plateau}$, it was initially estimated from the maximum smoothed hydration number and iteratively updated to the mean over the post-plateau interval until $t_{plateau}$ and $t_{end}$ converged. The vertical lines indicate $t_{plateau}$ and $t_{end}$ in panel (i) and $t_{plateau}$ in panel (ii); the resulting $\Delta x_{ref}$ values are annotated in panel (i).

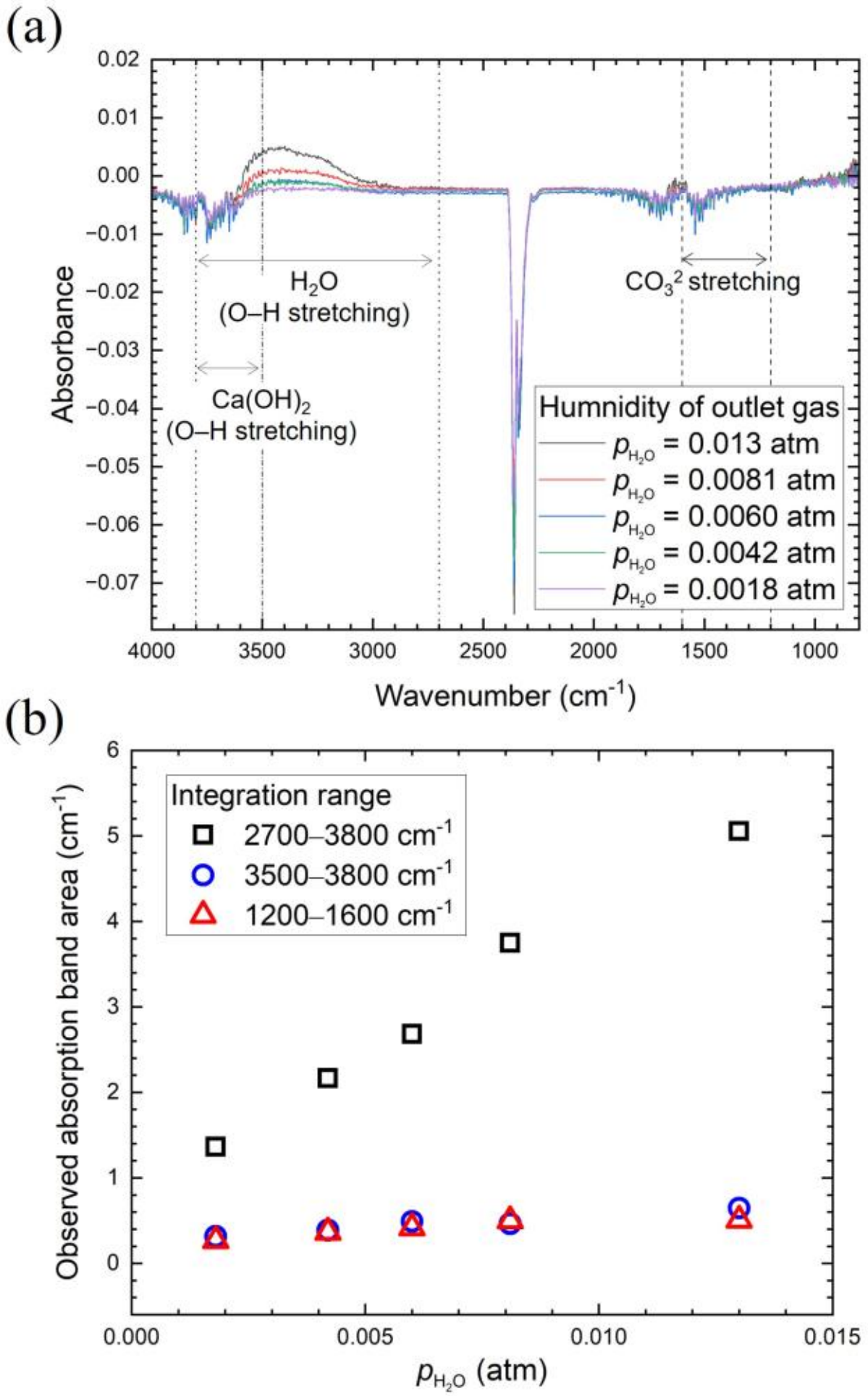

**Figure S5** Humidity dependence of the background spectra. (a) Background spectra measured using an Au-coated glass substrate without a sample at outlet $p_{H_2O}$ of 0.0018, 0.0042, 0.0060, 0.0081, 0.013 atm. These $p_{H_2O}$ values correspond to dry/humidified $N_2$ flow rates of 800/20, 650/150, 400/200, 200/200, and 0/200 ml mim$^{-1}$, respectively. Humidified $N_2$ was generated by bubbling $N_2$ through water at 10 °C. The double-headed arrows indicate the wavenumber ranges used to evaluate the water-related absorption band for $La_2(SO_4)_3$ and $CaSO_4$ (2700–3800 cm$^{-1}$), the hydroxyl-related absorption band for CaO (3500–3800 cm$^{-1}$), and the carbonate-related absorption band for CaO (1200–1600 cm$^{-1}$). (b) Absorbance areas in these wavenumber ranges as a function of outlet $p_{H_2O}$ values. The areas were calculated using the same baseline correction and integration procedure as for the sample spectra (section 3.3). Because the integration ranges include regions without distinct absorption bands, the measured areas represent the humidity-dependent background response within each range. Under the $p_{H_2O}$ conditions used for the in situ FTIR measurements, the background absorbance area in the 2700–3800 cm$^{-1}$ range was estimated to be 1–2 cm$^{-1}$. This indicates that the background contribution to the sample signals was small.

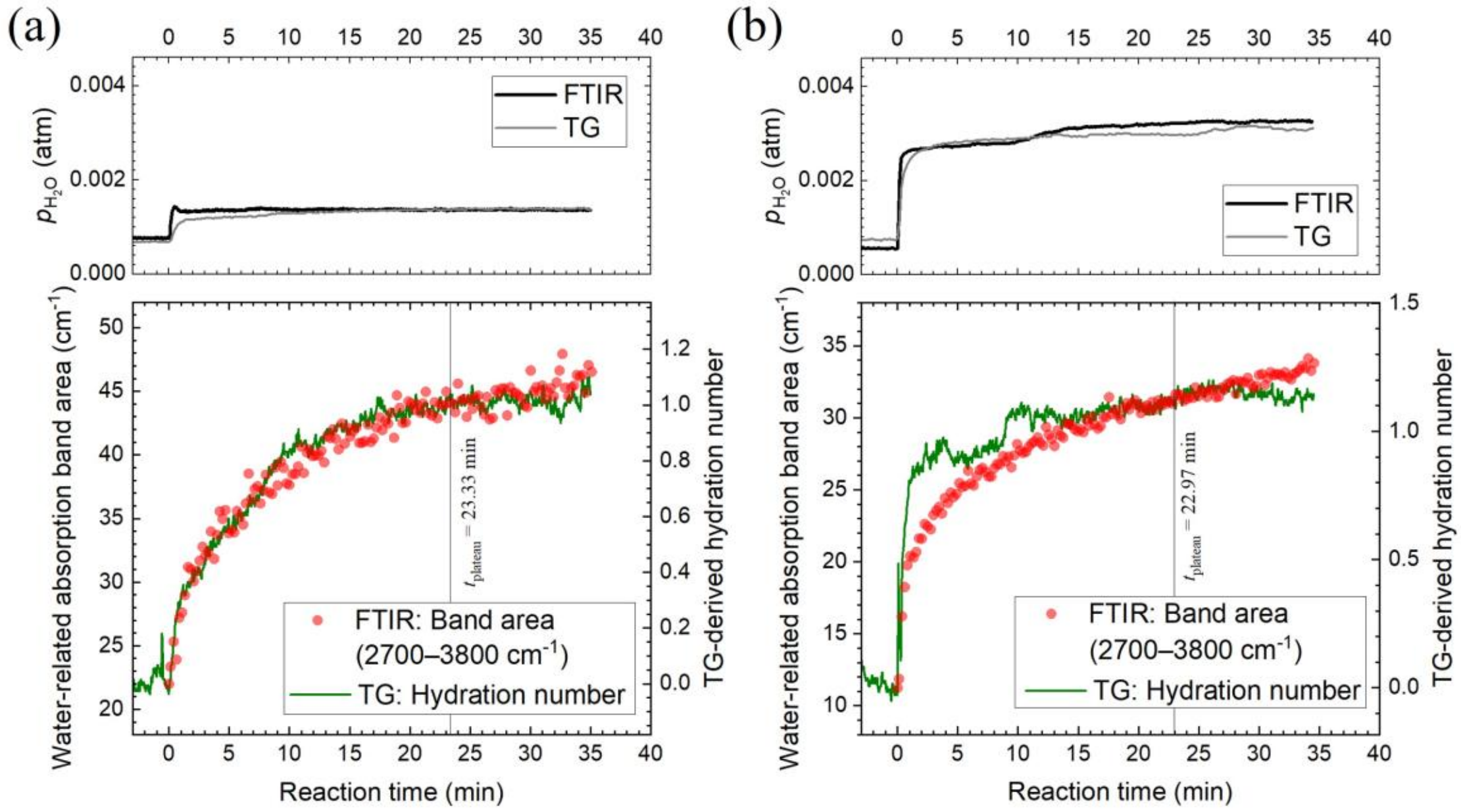

**Figure S6** Comparison of micro-FTIR and TG hydration profiles of $La_2(SO_4)_3$ under different $p_{H_2O}$. Measurements were performed under dry/humidified $N_2$ flow rates of (a) 800/20 and (b) 700/100 mL min$^{-1}$. The upper panels show the $p_{H_2O}$ profiles measured during the micro-FTIR and TG experiments. The lower panels compare the water-related absorption band area integrated over 2700–3800 cm$^{-1}$ with the TG-derived hydration number. Although the FTIR and TG profiles showed similar overall temporal trends, some deviation was observed, particularly at longer reaction times under the 700/100 mL min$^{-1}$ condition. The contribution of experimental factors, including drift in the FTIR signal, to this deviation cannot be excluded.

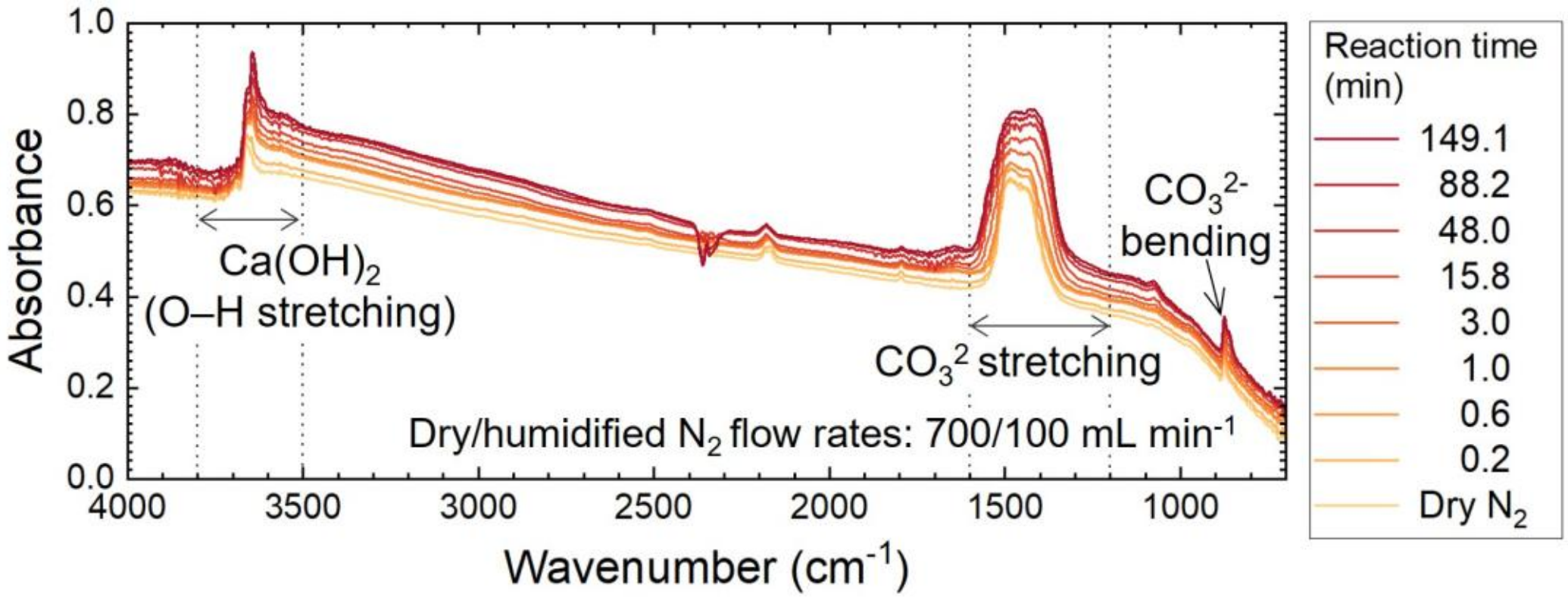

**Figure S7** In situ micro-FTIR spectra recorded at 30 °C during CaO hydration under dry/humidified $N_2$ flow rates of 700/100 mL min$^{-1}$. The humidified $N_2$ stream was generated by bubbling $N_2$ through water maintained at 10 °C. Selected spectra are shown from the initial dry $N_2$ atmosphere to a reaction time of 149.1 min. The spectrum labeled "Dry $N_2$" was acquired immediately before humidification. Reaction time was defined as the elapsed time after switching the gas flow from dry $N_2$ to the prescribed mixture of dry and humidified $N_2$. The hydroxyl- and carbonate-related absorption band area changes shown in Fig. 5(c) were calculated from this spectral dataset using integration ranges of 3500–3800 and 1200–1600 cm$^{-1}$, indicated by double-headed arrows, respectively. The sharp features near 2350 cm$^{-1}$ originate from atmospheric $CO_2$ in the optical path and are not sample-related.